\documentclass[twocolumn]{aastex631}

\usepackage{float}
\usepackage{hyperref}
\usepackage{lineno}

\begin{document}

\title{The ALMA-QUARKS Survey: Properties of Hot Molecular Fragments in the Massive Protocluster IRAS 17233–3606}

\author[0009-0009-8154-4205]{Li Chen}
\affiliation{School of Physics and Astronomy, Yunnan University, Kunming 650091, People’s Republic of China}
\correspondingauthor{Li Chen}
\email{leonchen0311@163.com}

\author[0000-0003-2302-0613]{Sheng-Li Qin}
\affiliation{School of Physics and Astronomy, Yunnan University, Kunming 650091, People’s Republic of China}

\author[0009-0004-6159-5375]{Dongting Yang}
\affiliation{Shanghai Astronomical Observatory, Chinese Academy of Sciences, 80 Nandan Road, Shanghai 200030, People's Republic of China}

\author[0000-0001-9822-7817]{Wenyu Jiao}
\affiliation{Shanghai Astronomical Observatory, Chinese Academy of Sciences, 80 Nandan Road, Shanghai 200030, People's Republic of China} 

\author[0000-0002-5286-2564]{Tie Liu}
\affiliation{Shanghai Astronomical Observatory, Chinese Academy of Sciences, 80 Nandan Road, Shanghai 200030, People's Republic of China}

\author[0000-0002-6622-8396]{Paul F. Goldsmith}
\affiliation{Jet Propulsion Laboratory, California Institute of Technology, 4800 Oak Grove Drive, Pasadena, CA 91109, USA}

\author[0009-0005-4295-5010]{Zhenying Zhang}
\affiliation{School of Physics and Astronomy, Yunnan University, Kunming 650091, People’s Republic of China}

\author[0000-0001-5710-6509]{Chuanshou Li}
\affiliation{School of Physics and Astronomy, Yunnan University, Kunming 650091, People’s Republic of China}

\author[0000-0001-5950-1932]{Fengwei Xu}
\affiliation{Max Planck Institute for Astronomy, Königstuhl 17, 69117, Heidelberg, Germany}

\author[0000-0002-8697-9808]{Sami Dib}
\affiliation{Max Planck Institute for Astronomy, Königstuhl 17, 69117, Heidelberg, Germany}

\author[0000-0002-8614-0025]{Shivani Gupta}
\affiliation{Indian Institute of Astrophysics, Koramangala II Block, Bangalore 560 034, India}
\affiliation{Pondicherry University, R.V. Nagar, Kalapet, 605014, Puducherry, India}

\author[0000-0002-4154-4309]{Xindi Tang}
\affiliation{Xinjiang Astronomical Observatory, Chinese Academy of Sciences, 830011 Urumqi, People’s Republic of China}

\author[0000-0001-5703-1420]{Yaping Peng}
\affiliation{Department of Physics, Faculty of Science, Kunming University of Science and Technology, Kunming 650500, People's Republic of China}

\author[0000-0001-9160-2944]{Mengyao Tang}
\affiliation{Institute of Astrophysics, School of Physics and Electronic Science, Chuxiong Normal University, Chuxiong 675000, People’s Republic of China}

\author[0000-0002-5809-4834]{Mika Juvela}
\affiliation{Department of Physics, P.O.Box 64, FI-00014, University of Helsinki,
Finland}

\author[0000-0003-4546-2623]{Di Li}
\affiliation{New Cornerstone Science Laboratory, Department of Astronomy, Tsinghua University, Beijing 100084, China}
\affiliation{Radio Astronomy National Key Laboratory, NAOC, Chinese Academy of Sciences, Beijing 100101, China}

\author[0000-0003-4546-2623]{Aiyuan Yang}
\affiliation{National Astronomical Observatories, Chinese Academy of Sciences, Beijing 100101, People's Republic of China}
\affiliation{Key Laboratory of Radio Astronomy and Technology, Chinese Academy of Sciences, A20 Datun Road, Chaoyang District, Beijing, 100101, People's Republic of China}

\author[0000-0002-5789-7504]{Meizhu Liu}
\affiliation{Center for Astrophysics, Guangzhou University, Guangzhou 510006, People’s Republic of China}

\author{Lingmin Zhen}
\affiliation{School of Physics and Astronomy, Yunnan University, Kunming 650091, People’s Republic of China}

\author[0000-0002-9875-7436]{James O. Chibueze}
\affiliation{UNISA Centre for Astrophysics and Space Sciences (UCASS),
College of Science, Engineering and Technology, University of South Africa, Cnr Christian de Wet Rd and Pioneer Avenue, Florida Park, 1709, Roodepoort, South Africa}
\affiliation{Department of Physics and Astronomy, Faculty of Physical Sciences, University of Nigeria, Carver Building, 1 University Road, Nsukka 410001, Nigeria}

\author[0000-0002-5310-4212]{L. Viktor T\'{o}th}
\affiliation{Department of Astronomy, E\"{o}tv\"{o}s Lor\'{a}nd University, P\'{a}zm\'{a}ny P\'{e}ter s\'{e}t\'{a}ny 1/A, H-1117, Budapest, Hungary }
\affiliation{Faculty of Science and Technology, University of Debrecen, H-4032 Debrecen, Hungary}

\author[0009-0003-6633-525X]{Ariful Hoque}
\affiliation{S. N. Bose National Centre for Basic Sciences, Block-JD, Sector-III, Salt Lake City, Kolkata 700106, India}

\author[0000-0003-2300-8200]{Amelia M.\ Stutz}
\affiliation{Departamento de Astronom\'{i}a, Universidad de Concepci\'{o}n,Casilla 160-C, Concepci\'{o}n, Chile}

\author[0000-0002-9574-8454]{Leonardo Bronfman}
\affiliation{Departamento de Astronomía, Universidad de Chile, Casilla 36-D, Santiago, Chile}

\author[0000-0001-7151-0882]{Swagat R. Das}
\affiliation{Departamento de Astronomia, Universidad de Chile, Las Condes, 7591245 Santiago, Chile}

\author[0000-0002-8586-6721]{Pablo Garc\'ia}
\affiliation{Chinese Academy of Sciences South America Center for Astronomy, National Astronomical Observatories, CAS, Beijing 100101, People's Republic of China}
\affiliation{Instituto de Astronomía, Universidad Católica del Norte, Av. Angamos 0610, Antofagasta, Chile}

\author[0000-0003-2412-7092]{Kee-Tae Kim}
\affiliation{Korea Astronomy and Space Science Institute, 776 Daedeokdae-ro, Yuseong-gu, Daejeon 34055, Republic of Korea}
\affiliation{University of Science and Technology, Korea (UST), 217 Gajeong-ro, Yuseong-gu, Daejeon 34113, Republic of Korea}

\author[0000-0002-3179-6334]{Chang Won Lee}
\affiliation{Korea Astronomy and Space Science Institute, 776 Daedeokdaero,  Yuseong-gu, Daejeon 34055, Republic of Korea}
\affiliation{University of Science and Technology, Korea (UST), 217  Gajeong-ro, Yuseong-gu, Daejeon 34113, Republic of Korea}

\author[0000-0001-7866-2686]{Jihye Hwang}
\affiliation{Korea Astronomy and Space Science Institute, 776 Daedeokdaero,  Yuseong-gu, Daejeon 34055, Republic of Korea}

\author[0009-0000-9090-9960]{Jiahang Zou}
\affiliation{School of Physics and Astronomy, Yunnan University, Kunming 650091, People’s Republic of China}

\author{Yongqi Guo}
\affiliation{School of Physics and Astronomy, Yunnan University, Kunming 650091, People’s Republic of China}

\author{Zhiping Kou}
\affiliation{State Key Laboratory of Radio Astronomy and Technology, Xinjiang Astronomical Observatory, CAS, 150 Science 1-Street, Urumqi, Xinjiang 830011, People’s Republic of China}
\affiliation{University of Chinese Academy of Sciences, Beijing 100049, People’s Republic of China}


\begin{abstract}
To investigate the physical mechanisms of fragmentation within the hot molecular core of the massive protocluster IRAS 17233-3606 (G351.78-0.54), we carried out a detailed analysis of continuum and lines, using the ALMA Band 3 data from the ATOMS survey and Band 6 data from the QUARKS survey. The low-resolution 3 mm data reveal a massive hot core MM1 with a mass of $\sim$81.3 M$_\odot$, and a prominent ultracompact (UC) H{\sc ii} region MM2, while the high-resolution data resolve MM1 into 11 hot molecular fragments (HMFs). These HMFs exhibit hot (T$_{\rm rot}$ = 100–310 K) CH$_3$CN and CH$_3$OH emission and high column densities (N$\rm_{H_2}$ $>$ 10$^{23}$ cm$^{-2}$), indicating their potential to form massive stars. 
Based on outflows, masers, H{\sc ii} regions, and f[CH$_3$CN/CH$_3$OH] abundance ratios, the evolutionary sequences of the 11 HMFs are categorized as phases I to IV. 
The mean minimum-spanning tree (MST) separation ($\sim$$1.8\times 10^3$ au) of the HMFs is nearly half of the thermal Jeans length ($\sim$$3.3\times 10^3$ au). Together with the $Q$ parameter $Q = 0.77$ and virial parameter $\alpha_{\rm vir}=0.84$ of MM1, these results suggest an evolutionary scenario in which fragmentation is initially driven by thermal instability, followed by global gravitational contraction and growth through active accretion.
Meanwhile, feedback from the B2-type zero-age main-sequence (ZAMS) star and the UC H{\sc ii} region significantly influence the morphology and chemical properties of MM1 and MM2. 
This heterogeneity highlights the role of diverse physical processes taking place in high-mass protoclusters.

\end{abstract}


\section{Introduction} \label{sec:intro}

High-mass star formation (HMSF) is a complex process that occurs on short timescales ($\lesssim$ 10$^{6}$ years) and is driven by the interplay of multiple physical processes \citep{1998MNRAS.295..691B, 2007prpl.conf..165B, 2007ARA&A..45..565M, 2010MNRAS.405..401D, 2014prpl.conf..149T, 2018ARA&A..56...41M, 2025A&A...695A..51B}. 
Massive stars are thought to form preferentially in clustered environments embedded in parsec-scale molecular clumps \citep{2003ARA&A..41...57L,2007ARA&A..45..481Z}. Within clumps, density fluctuations induced by thermal Jeans instability or turbulent trigger fragmentation, produce a population of dense cores with characteristic sizes of $\sim$0.1 pc \citep{2009ApJ...696..268Z, 2011A&A...530A.118P,2015ApJ...804..141Z,2018A&A...617A.100B,2019ApJ...886..102S,2023MNRAS.526.2278A,2024ApJ...966..171M,2025ApJS..280...33Y}. These dense cores may subsequently undergo further fragmentation and then form stars. In such protocluster environments, understanding how massive protostars assemble, rapidly accrete material, and undergo early evolution in chemical abundance and structural morphology remains a pivotal challenge \citep{2019ARA&A..57..227K, 2020ApJ...900...82P, 2023A&A...674A.160G, 2023ApJ...959...88D, 2024ApJS..270....9X, 2025A&A...696A...7L}. The analysis of these processes is of great significance for elucidating the underlying mechanisms of high-mass star formation, the early structures of protoclusters, and the origins and evolution of their chemical environments \citep{1998ARA&A..36..317V, 1999ARA&A..37..311E, 2005ASSL..324...87T, 2018ASPC..517..249B, 2024A&A...686A.252B}.

High-resolution observations from the Atacama Large Millimeter/submillimeter Array (ALMA) have provided key insights into dense core evolution and fragmentation. The ALMA observations have not only revealed the coexistence of dense cores spanning different evolutionary stages, including starless cores, hot molecular cores (HMCs), and ultra-compact H{\sc ii} regions \citep{2025ApJS..280...33Y}, but also resolved intricate substructures within individual cores \citep{2025ApJ...994..233Y,2026ApJ..1003...37F,2026ApJ...997..340M}. HMCs are typically defined as compact regions ($\leq$ 0.1 pc) with high temperatures ($\ge$ 100 K), high density (n$\rm _{H_2}$ $\gtrsim$ 10$^6$ cm$^{-3}$), and rich in complex organic molecules (COMs) \citep{1999PASP..111.1049G, 1999ApJ...525..808O, 2000prpl.conf..299K, 2005IAUS..227...59C}. With the increase in observational resolution, hot molecular fragments (HMFs) were found to be the internal substructures within HMCs \citep{2026ApJ...997..340M}. It appears that HMCs and HMFs refer to the same physical entities, which are only observed at different scales.

During the HMC/HMF phase, as the embedded (proto-)stars and their accretion envelopes heat the surrounding material, molecules originally frozen on dust surfaces (ice mantles) undergo desorption (through thermal desorption or triggered by shocks).  Once released into the gas phase, these species initiate complex chemical reactions that produce large quantities of COMs \citep{2004MNRAS.354.1141V, 2006A&A...457..927G, 2009ARA&A..47..427H, 2013ApJ...765...60G, 2022ApJS..259....1G, 2022A&A...665A..96B}. This phase may also generate observable molecular outflows. Consequently, the HMC/HMF serve as an important window for studying the physical and chemical conditions of early high-mass star formation through (sub)millimeter molecular line observations \citep{2018IAUS..332....3V, 2022MNRAS.512.4419P, 2024ApJ...962...13C}. 
Based on the spatial distributions of masers, outflows, and H{\sc ii} regions, the evolutionary sequence of hot cores can be divided into six distinct phases \citep{2004A&A...417..615C, 2017A&A...604A..60B, 2025A&A...693A.160M}.
However, the physical and chemical evolution of hot cores, especially during the earliest phases (I–III), still lack sufficient observational evidence to characterize the underlying physical processes and chemical properties in these early stages.

IRAS 17233–3606 (also known as G351.78-0.54) is an ideal target for studying high-mass star formation, which is located at coordinates of R.A. $=17^{\rm h}26^{\rm m}42^{\rm s}.800$ and Dec. $=-36^\circ 09' 16''.80$. It is characterized by its high luminosity (L$\rm_{bol}$ $\approx$ 4$\times$10$^4$ L$_\odot$, \citealt{2018MNRAS.473.1059U}), close distance of 1.32 kpc (\citealt{2024RAA....24b5009L}), active maser emission (e.g., \citealt{1980IAUC.3509....2C, 1983AuJPh..36..361C, 1989A&A...213..339F, 1991ApJ...380L..75M, 1998MNRAS.301..640W, 2005ApJS..160..220F}), rich in hot core chemistry\citep{2022MNRAS.511.3463Q, 2023ApJ...950...57T, 2025ApJ...983...37S, 2025A&A...697A.190L}, and multiple strong outflows \citep{2008A&A...485..167L, 2009A&A...507.1443L, 2011A&A...530A..12L, 2013A&A...554A..35L, 2025ApJ...987..197H}. Previous studies with different resolutions have resolved filamentary inflows on large scales \citep{2021MNRAS.505..726R}, disk-like rotation \citep{2017A&A...603A..10B}, as well as several fragments around the central high-mass source on smaller scales \citep{2019A&A...621A.122B, 2025A&A...695A..51B}. However, the dense core population in the central region and its evolutionary state remain unresolved.

This paper presents the observational results from ALMA-ATOMS Band 3 ($\sim1\farcs7$, $\sim$$2.2\times 10^3$ au) and ALMA-QUARKS Band 6 ($\sim0\farcs3$, $\sim$$4.0\times 10^2$ au) observations, aimed at exploring the star formation activity in IRAS 17233-3606, with an emphasis on resolving the internal structures and dynamical processes of the massive protocluster. This paper is structured as follows: observations are described in Section \ref{obs}; the results are presented in Section \ref{res}; the discussion is followed in Section \ref{dis}; and finally, the main conclusions are given in Section \ref{con}.

\section{ALMA Observations}
\label{obs}

IRAS 17233-3606 was observed as part of the ALMA Three-millimeter Observations of Massive Star-forming regions (ATOMS) survey (PI: Tie Liu; Project ID: 2019.1.00685.S; \citealt{2020MNRAS.496.2790L}), with both the ALMA 12 m array and the 7 m Atacama Compact Array (ACA), using eight spectral windows. 
Eight spectral windows (SPWs) include six high-resolution SPWs and two broadband SPWs with a 1875 MHz bandwidth. The sensitivity reaches 12 mJy beam$^{-1}$ per 0.062 MHz channel ($\sim$0.2-0.4 km s$^{-1}$) for six SPWs, and 3.6 mJy beam$^{-1}$ per 0.488 MHz channel ($\sim$2.9 km s$^{-1}$) for two broadband SPWs.
The on-source integration totaled ~3 min and ~8 min for the 12 m array and ACA, respectively. The 12 m data, taken on 2019 November 3, have a synthesized beam of $\sim2\farcs17\times1\farcs85$ and a maximum recoverable scale (MRS) of $\sim20\farcs1$.
The ACA observations, obtained on 2019 November 19, have a resolution of $\sim13\farcs3$ and an MRS of $\sim76\farcs2$, with an rms of 83 mJy beam$^{-1}$ per 0.122 MHz channel. Calibration and imaging were carried out using CASA version 5.6 \citep{2007ASPC..376..127M}. The ACA and 12 m datasets were processed separately and subsequently combined during imaging to recover extended emission filtered out by the 12 m array alone.

Follow-up observations at 1.3 mm were carried out as part of the Querying Underlying Mechanisms of Massive Star Formation with the ALMA-Resolved Gas Kinematics and Structures (QUARKS) survey (PIs: Lei Zhu, Guido Garay, and Tie Liu; Project ID: 2021.1.00095.S; \citealt{2024RAA....24b5009L,2024RAA....24f5011X}). The observations were conducted with the ALMA 12 m array (configurations C2 + C5) and the ACA in four spectral windows, providing a total bandwidth of 7.5 GHz. Each broadband spectral window has a bandwidth of 1875 MHz, yielding a sensitivity of $\sim5$ mJy beam$^{-1}$ per 0.488 MHz channel ($\sim0.67$ km s$^{-1}$). The combined data achieve a synthesized beam of $\sim0\farcs39\times0\farcs27$ for this source. Details of the data reduction procedures can be found in \citet{2024RAA....24b5009L}. High-resolution Band 6 observations provide a more detailed view of the molecular outflows and allow us to identify their driving sources through the improved spatial correlation between the continuum peaks and the outflow lobes.

\section{RESULTS}
\label{res}

\subsection{HMFs Identification}
\label{DHC}

\begin{figure*}[ht!]
\centering
\includegraphics[width=\linewidth]{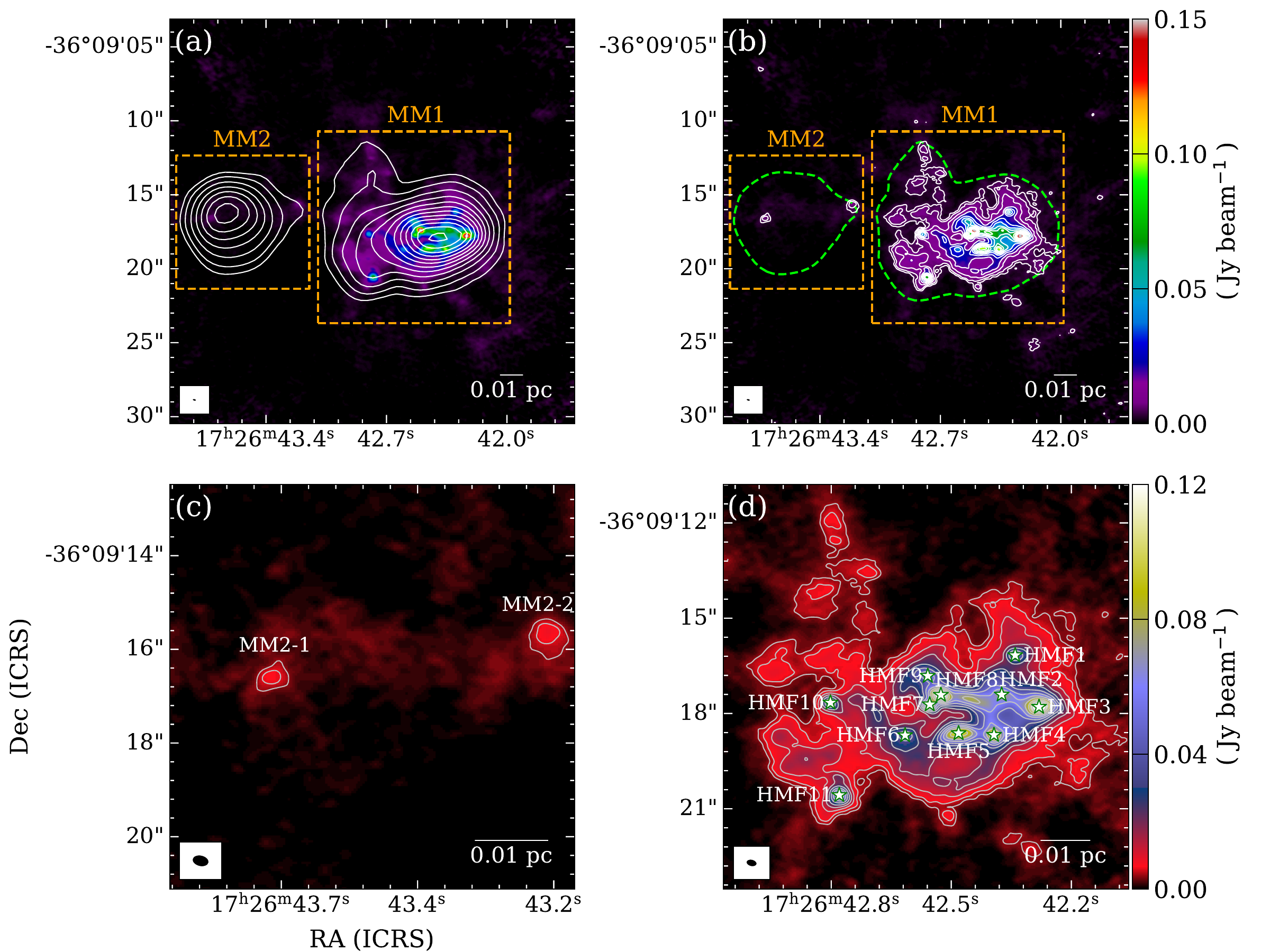}
\caption{
ALMA Band 3 and 6 continuum emission of IRAS 17233-3606. 
Panel (a) shows the Band 6 continuum emission with Band 3 continuum emission overplotted as white contours of $\rm D = 4\times N^{p} + 8$, where $N=12$ is the number of contours used and $\rm p = \log(V_{\rm max}/(4\times \rm rms))/\log(N-1)$ is the power index, with rms = $1\ \rm mJy\ beam^{-1}$. The orange dashed squares labeled MM1 (right) and MM2 (left) identify a hot core region and an UC H{\sc ii} region extracted in Band 3 from \cite{2022MNRAS.511.3463Q}, respectively. 
Panel (b) shows the white contours of Band 6 continuum emission overplotted on the same color map as panel (a). The white contour levels follow the same rule in panel (a) but use an rms of $0.5\ \rm mJy\ beam^{-1}$. The dashed green contour outlines the $7.8\ \rm mJy\ beam^{-1}$ detection level of Band 3 continuum emission. 
Panel (c) shows a zoom toward the UC H{\sc ii} region MM2, with contours at 4.0 and 6.0 $\rm mJy\ beam^{-1}$, the same as in panel (b). 
Panel (d) shows a zoom toward the hot core region MM1, with contours the same as in panel (b). The white stars mark the HMFs identified in the Band 6 observations.
The synthesized beams are shown in the bottom left corners and the scale bar are shown in the bottom right corners for each panel.}
\label{zoomin}
\end{figure*}

Figure~\ref{zoomin} presents the ALMA Band 3 and Band 6 continuum observations of IRAS 17233-3606. The Band 3 continuum emission reveals a hot core region MM1 identified via molecular line emission from C$_2$H$_5$CN, CH$_3$OCHO, and CH$_3$OH \citep{2022MNRAS.511.3463Q}, and a UC H{\sc ii} region MM2 identified by H40$\alpha$ emission. Higher-resolution Band 6 observations ($\sim0\farcs3$, or $\sim$400 au) resolve both MM1 and MM2 regions into continuum substructures. 
The hot core MM1 is resolved into multiple separate HMFs in the high-resolution 1.3 mm emission. MM2 exhibits strong 3 mm emission but weak 1.3 mm emission, consistent with a dominant free-free contribution.

\begin{deluxetable*}{cllcccccc}[ht!]
\tablewidth{0pt}
\tablecaption{Parameters of the HMFs observed in Band 6.\label{tab:cores}}
\tablehead{
\colhead{ID} & \colhead{RA} & \colhead{DEC} & \colhead{$\theta_{\rm maj}^{dec} \times \theta_{\rm min}^{dec}$} & \colhead{PA} & \colhead{S$^{peak}_{\nu}$} & \colhead{S$^{int}_{\nu}$} & \colhead{Mass} & \colhead{N$\rm _{H_2}$} \\
\colhead{} & \colhead{(h m s)} & \colhead{($^{\circ}$~$^\prime$~$^{\prime\prime}$)} & \colhead{($^{\prime\prime}$$\times$$^{\prime\prime}$)} & \colhead{($^{\circ}$)} & \colhead{(mJy/beam)} & \colhead{(mJy)} & \colhead{($\rm M_\odot$)} & \colhead{(cm$^{-2}$)} \\
\colhead{(1)} & \colhead{(2)} & \colhead{(3)} & \colhead{(4)} & \colhead{(5)} & \colhead{(6)} & \colhead{(7)} & \colhead{(8)} & \colhead{(9)} 
}
\startdata
HMF1 & 17:26:42.336 & $-$36:09:16.180 & 0.531$\times$0.427 & 97$\pm$12 & 48$\pm$2 & 157$\pm$6 & 0.68$\pm$0.04 & (1.04$\pm$0.06)$\times$10$^{24}$ \\
HMF2 & 17:26:42.373 & $-$36:09:17.437 & 1.074$\times$0.723 & 98$\pm$3 & 75$\pm$1 & 629$\pm$12 & 1.93$\pm$0.12 & (8.62$\pm$0.55)$\times$10$^{23}$ \\
HMF3 & 17:26:42.276 & $-$36:09:17.792 & 0.624$\times$0.412 & 101$\pm$5 & 164$\pm$5 & 574$\pm$24 & 1.98$\pm$0.08 & (2.68$\pm$0.11)$\times$10$^{24}$ \\
HMF4 & 17:26:42.394 & $-$36:09:18.682 & 0.542$\times$0.487 & 75$\pm$20 & 102$\pm$3 & 361$\pm$12 & 1.07$\pm$0.06 & (1.41$\pm$0.07)$\times$10$^{24}$ \\
HMF5 & 17:26:42.482 & $-$36:09:18.664 & 1.196$\times$0.524 & 97$\pm$2 & 99$\pm$1 & 721$\pm$12 & 1.62$\pm$0.07 & (8.95$\pm$0.36)$\times$10$^{23}$ \\
HMF6 & 17:26:42.624 & $-$36:09:18.712 & 1.016$\times$0.634 & 88$\pm$1 & 40$\pm$1 & 286$\pm$8 & 0.92$\pm$0.04 & (4.93$\pm$0.24)$\times$10$^{23}$ \\
HMF7 & 17:26:42.556 & $-$36:09:17.690 & 0.446$\times$0.280 & 122$\pm$19 & 93$\pm$6 & 214$\pm$20 & 0.65$\pm$0.04 & (1.81$\pm$0.12)$\times$10$^{24}$ \\
HMF8 & 17:26:42.534 & $-$36:09:17.442 & 0.680$\times$0.433 & 112$\pm$23 & 142$\pm$15 & 547$\pm$72 & 1.50$\pm$0.08 & (1.76$\pm$0.09)$\times$10$^{24}$ \\
HMF9 & 17:26:42.565 & $-$36:09:16.828 & 0.844$\times$0.520 & 65$\pm$1 & 55$\pm$1 & 288$\pm$4 & 0.82$\pm$0.05 & (6.49$\pm$0.41)$\times$10$^{23}$ \\
HMF10 & 17:26:42.818 & $-$36:09:17.678 & 0.380$\times$0.361 & 75$\pm$3 & 47$\pm$1 & 97$\pm$2 & 0.70$\pm$0.15 & (2.23$\pm$0.47)$\times$10$^{24}$ \\
HMF11 & 17:26:42.795 & $-$36:09:20.570 & 0.392$\times$0.363 & 170$\pm$44 & 72$\pm$2 & 162$\pm$6 & 0.62$\pm$0.03 & (1.71$\pm$0.09)$\times$10$^{24}$ \\
\enddata
\tablecomments{The 11 HMFs are listed in the table, with their detailed physical parameters given in columns (2)–(7), derived from two-dimensional Gaussian fits in CASA. The HMF masses and H$_2$ column densities are presented in columns (8) and (9), respectively, with uncertainties propagated from errors in flux, gas–dust ratio, and dust temperature.}
\end{deluxetable*}

Using the core extraction algorithm developed by \citet{2024ApJ...962...13C}, we identified 11 HMFs from the 1.3~mm continuum map.
Firstly, dozens of dense cores were selected as HMF candidates by searching for local intensity peaks in the continuum emission.
We then extracted the core-averaged spectra, modeling the line emission under the assumption of local thermodynamic equilibrium (LTE) using the eXtended CASA Line Analysis Software Suite (XCLASS; \citealt{2017A&A...598A...7M}). Ultimately, 11 HMFs were confirmed as rich in COMs based on the detection of CH$_3$CN and CH$_3$OH emission with rotational temperatures exceeding 100 K.
The observational properties of these HMFs are derived via 2D Gaussian fitting in CASA and are listed in Table~\ref{tab:cores}, including coordinate position, size, position angle (PA), peak intensity and integrated flux density for each HMF. 

\begin{figure*}[ht!]
\centering
\includegraphics[angle=270,width=0.95\linewidth]{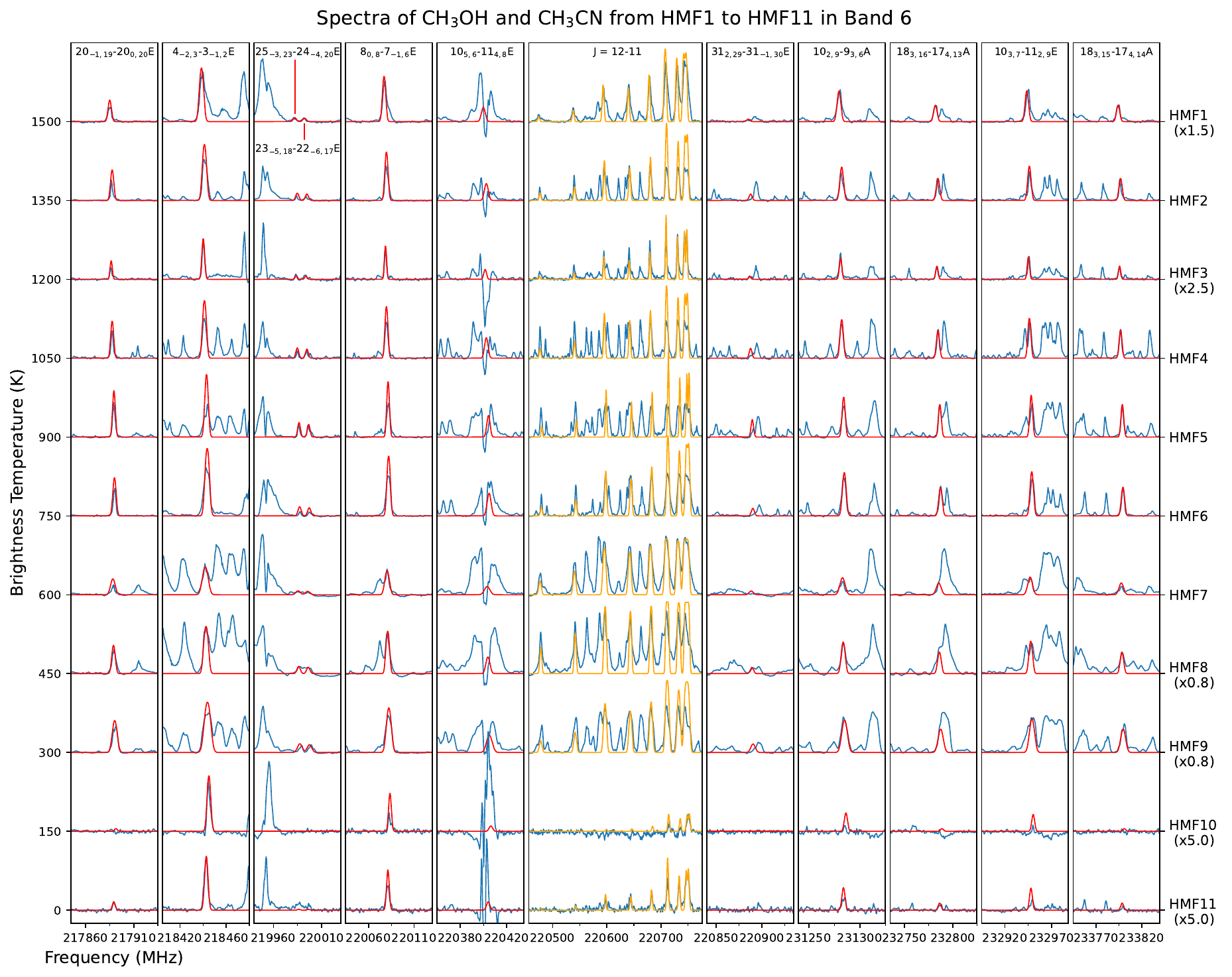}
\caption{Best-fit CH$_3$OH (red) and CH$_3$CN (orange) spectra toward the 11 HMFs (HMF1–HMF11), superimposed on observed core-averaged spectra (blue). The corresponding quantum numbers for each transition are listed at the top of each panel. A vertical offset of 150 K and intensity scaling (factors are indicated on the right) are applied for clarity. Features in the observed spectra (blue) not accounted for by the fit correspond to emission from other molecules.}
\label{spectra_MM1}
\end{figure*}

Table~\ref{tab:mol_data} presents the fitted rotational temperatures, column densities, the full width at half maximum (FWHM) and velocity offset of CH$_3$CN and CH$_3$OH and their relative abundance.
Figure~\ref{spectra_MM1} shows the best-fit spectra of CH$_3$OH and CH$_3$CN.
For CH$_3$OH, due to spectral blending with CH$_3$COCH$_3$ for the $23_{-5,18}-22_{-6,17}$E transition, with $^{13}$CO and CH$_3$CN for the $10_{5,6}-11_{4,8}$E transition, with CH$_3$OCHO for the $31_{2,9}-31_{-1,30}$E transition, and with C$_2$H$_5$CN for the $18_{3,16}-17_{4,13}$A transition, we exclude these four transitions and utilize the remaining seven for XCLASS modeling.
However, for CH$_3$CN, since the emission lines of low energy levels are typically optically thick, resulting in broadened line profiles and flattened peaks that distort the line shapes. Besides, the high density of rotational transitions within a narrow frequency range results in severe spectral blending. To mitigate these issues, we implemented a selective fitting strategy, prioritizing the fitting of isolated, optically thin, or moderately thick transitions that cover a wide range of excitation energies.
Finally, the consistency between the temperature distributions of CH$_3$OH and CH$_3$CN suggests that the temperature estimates are reliable. 

\subsection{Parameters of HMFs}

The masses of the HMFs are estimated from the 1.3 mm continuum map using the formula of \citet{1983QJRAS..24..267H}:
\begin{equation}
 \label{eq:core_mass}
\rm M_{\text{HMF}} = \frac{D^2 S_\nu^{int} \eta}{\kappa_\nu B_\nu (T_d)},
\end{equation}
where $D$ is the distance of 1.32 kpc to the Sun, $S_{\nu}^{\rm int}$ is the integrated continuum flux, $\eta$ = 111.9$\pm$1.2 is the gas-to-dust ratio, derived for galactocentric distance R$\rm_{GC}$ = 7.0 kpc using the relation $\rm \log(\eta) = \left(0.087\pm0.007\right) R_{\text{GC}} + \left(1.44\pm0.03\right)$ from \cite{2017A&A...606L..12G}, $\kappa_\nu$ is the dust absorption coefficient which for molecular cloud cores was interpolated to be 0.899 cm$^2$ g$^{-1}$ at 1.3 mm \citep{1994A&A...291..943O}, and $\rm B_\nu (T_d)$ is the Planck function at the dust temperature $\rm T_d$, which is set to the CH$_3$CN rotational temperature assuming gas-dust thermal equilibrium via efficient collisional coupling at high densities \citep{2001ApJ...557..736G}.
To assess the reliability of the calculated HMF masses, we derive the mean optical depth of the 1.3 mm continuum using the following equation \citep{2010ApJ...723.1665F,2021A&A...648A..66G}:
\begin{equation}
\rm \textit{$\tau_{\nu}$}=-ln[1-\frac{{S_\nu^{int}}}{\Omega B_{\nu}(T_d)}]
\end{equation}
where $\Omega$ is the solid angle subtended by the source. The derived $\tau_{\nu}$ toward the 11 HMFs are in the range of 0.019 to 0.106 (only HMF3 exceeds 0.1), with a mean value of 0.053. Therefore, optically thin assumption of dust continuum at 1.3 mm is reasonable and the calculated core masses are reliable.

\begin{deluxetable*}{llccc|lccc}[ht!]
\tablewidth{0pt}
\tablecaption{Parameters of CH$_3$CN and CH$_3$OH observed in Band 6.\label{tab:mol_data}}
\tablehead{
\colhead{} 
& \multicolumn{4}{c}{CH$_3$CN} 
& \multicolumn{4}{c}{CH$_3$OH} \\
\hline
\colhead{ID} & \colhead{T$\rm _{rot}$ } & \colhead{N} & \colhead{FWHM} & \colhead{V$\rm_{off}$} & \colhead{T$\rm _{rot}$ } & \colhead{N} & \colhead{FWHM} & \colhead{V$\rm_{off}$} \\
\colhead{} & \colhead{(K)} & \colhead{(cm$^{-2}$)} & \colhead{(km s$^{-1}$)} & \colhead{(km s$^{-1}$)} & \colhead{(K)} & \colhead{(cm$^{-2}$)} & \colhead{(km s$^{-1}$)} &  \colhead{(km s$^{-1}$)} \\
\colhead{(1)} & \colhead{(2)} & \colhead{(3)} & \colhead{(4)} & \colhead{(5)} & \colhead{(6)} & \colhead{(7)} & \colhead{(8)}  & \colhead{(9)} 
}
\startdata
HMF1 & 163$\pm$9 & (2.2$\pm$0.2)$\times$10$^{16}$ & 5.0$\pm$0.3 & 1.8$\pm$0.1 & 195$\pm$4 & (1.0$\pm$0.1)$\times$10$^{18}$ & 5.2$\pm$0.1 & 2.0$\pm$0.1 \\
HMF2 & 228$\pm$14 & (2.4$\pm$0.2)$\times$10$^{16}$ & 5.1$\pm$0.2 & -1.5$\pm$0.1 & 231$\pm$13 & (1.5$\pm$0.2)$\times$10$^{18}$ & 5.1$\pm$0.1 & -1.5$\pm$0.1 \\
HMF3 & 203$\pm$8 & (4.1$\pm$0.2)$\times$10$^{15}$ & 3.4$\pm$0.2 & -0.2$\pm$0.1 & 226$\pm$14 & (3.0$\pm$1.0)$\times$10$^{17}$ & 3.6$\pm$0.1 & -0.1$\pm$0.1 \\
HMF4 & 236$\pm$12 & (3.6$\pm$0.7)$\times$10$^{16}$ & 4.5$\pm$0.2 & -1.5$\pm$0.1 & 244$\pm$3 & (2.3$\pm$0.2)$\times$10$^{18}$ & 4.5$\pm$0.2 & -1.5$\pm$0.1 \\
HMF5 & 310$\pm$12 & (2.2$\pm$0.2)$\times$10$^{16}$ & 3.6$\pm$0.2 & -5.9$\pm$0.1 & 288$\pm$24 & (2.1$\pm$0.3)$\times$10$^{18}$ & 4.3$\pm$0.2 & -4.0$\pm$0.1 \\
HMF6 & 219$\pm$10 & (3.2$\pm$0.5)$\times$10$^{16}$ & 5.5$\pm$0.3 & -4.8$\pm$0.1 & 222$\pm$20 & (2.1$\pm$0.1)$\times$10$^{18}$ & 5.1$\pm$0.3 & -4.8$\pm$0.1 \\
HMF7 & 230$\pm$15 & (1.2$\pm$0.3)$\times$10$^{17}$ & 7.0$\pm$0.2 & -1.5$\pm$0.1 & 243$\pm$8 & (1.9$\pm$0.2)$\times$10$^{18}$ & 7.2$\pm$0.1 & -2.5$\pm$0.1 \\
HMF8 & 255$\pm$13 & (1.5$\pm$0.2)$\times$10$^{17}$ & 5.3$\pm$0.2 & -3.3$\pm$0.1 & 228$\pm$3 & (2.5$\pm$0.1)$\times$10$^{18}$ & 5.3$\pm$0.1 & -3.4$\pm$0.2 \\
HMF9 & 245$\pm$15 & (7.9$\pm$0.2)$\times$10$^{16}$ & 7.5$\pm$0.3 & -3.1$\pm$0.1 & 247$\pm$26 & (3.4$\pm$0.3)$\times$10$^{18}$ & 7.5$\pm$0.3 & -5.1$\pm$0.1 \\
HMF10 & 100$\pm$20 & (5.6$\pm$1.0)$\times$10$^{14}$ & 4.6$\pm$0.3 & -6.8$\pm$0.1 & 104$\pm$16 & (1.1$\pm$0.1)$\times$10$^{17}$ & 4.6$\pm$0.2 & -6.8$\pm$0.1 \\
HMF11 & 183$\pm$9 & (2.4$\pm$0.2)$\times$10$^{15}$ & 4.7$\pm$0.1 & -4.2$\pm$0.1 & 146$\pm$6 & (1.6$\pm$0.4)$\times$10$^{17}$ & 4.5$\pm$0.2 & -3.7$\pm$0.2 \\
\enddata
\tablecomments{The fitted rotational temperatures, column densities, full width at half maximum (FWHM) and velocity offset for CH$_3$CN and CH$_3$OH are given in columns (2)–(5) and (6)–(9), respectively.}
\end{deluxetable*}

Previous studies have shown that the formation of massive stars typically requires high column densities (e.g., reaching 1 g cm$^{-2}$ or $\rm N_{H_2}$ $\gtrsim$ 10$^{23}$ cm$^{-2}$) \citep{2003ApJ...585..850M, 2008Natur.451.1082K, 2010ApJ...723L...7K}. For the 11 HMFs, the source-averaged H$_2$ column density (N$_{\rm H_2}$) is derived as \citep{2010ApJ...723.1665F, 2019A&A...628A..27B}:
\begin{equation}
\label{eq:h2_cd}
\rm N_{\text{$\rm H_2$}} = \frac{S_\nu^{int} \eta}{\mu_{H_2} m_H \Omega  \kappa_\nu B_\nu (T_d)},
\end{equation}
where $\rm\mu_{H_2}\approx2.8$ is the mean particle weight per $\rm H_2$ molecule \citep{2008A&A...487..993K}, $\rm m_H$ is the mass of a hydrogen atom, $\Omega$ is the solid angle covered by the source. The resultant N$\rm_{H_2}$ ranges from 4.93$\times$10$^{23}$ to 2.68$\times$10$^{24}$ cm$^{-2}$, which supports the potential for massive star formation in MM1.

\subsection{Star-forming Activities}

\begin{figure*}
\centering
\includegraphics[width=\linewidth]{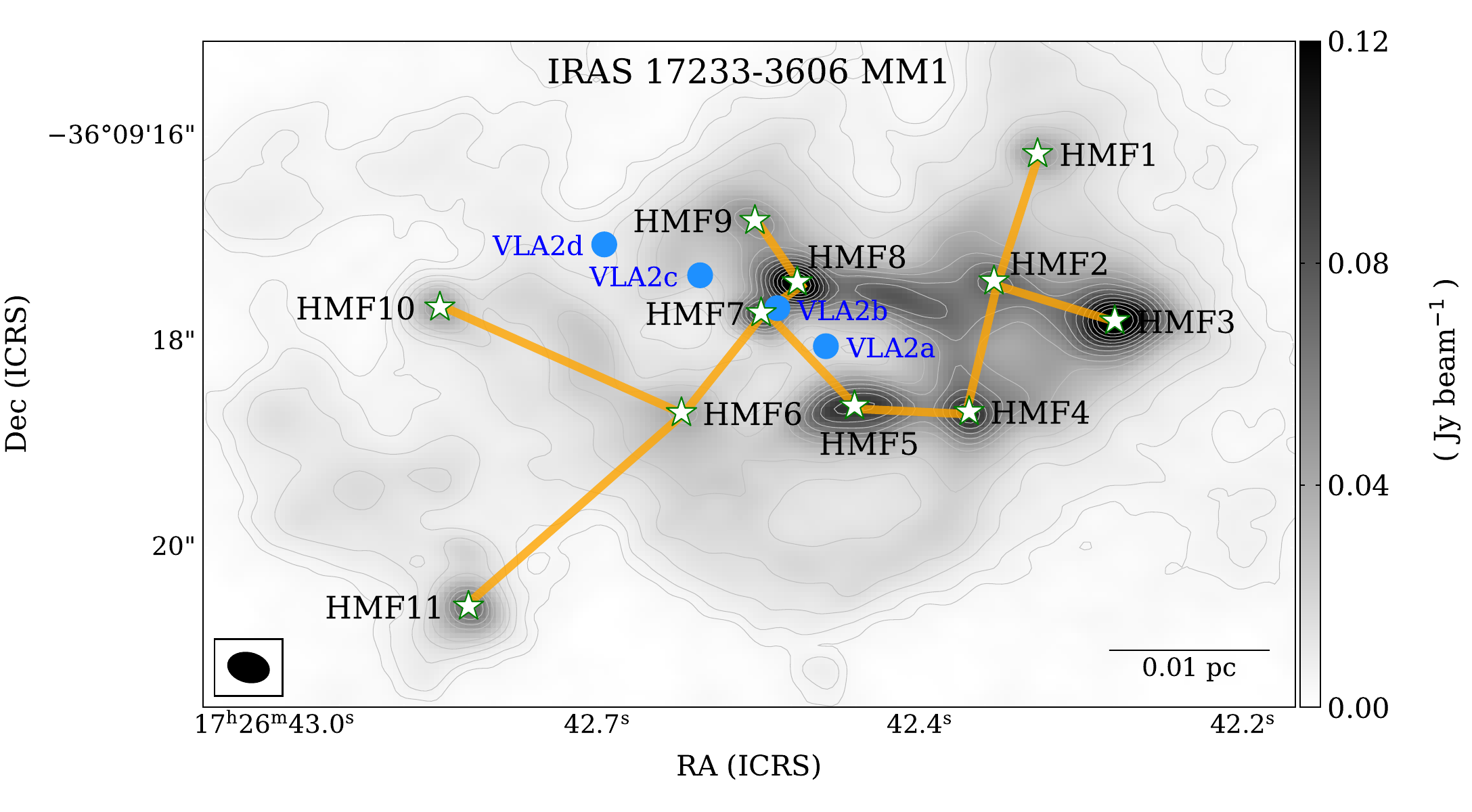}
\caption{Spatial distribution of HMFs in IRAS 17233-3606 MM1. The background displays the ALMA 1.3 mm continuum emission in grayscale. Gray contours are overlaid, following the same power-law progression as Figure~\ref{zoomin}. The white stars mark the locations of the HMFs identified by ALMA 1.3 mm observation, while the blue dots indicate the positions of the HC H{\sc ii} regions identified by VLA 1.3 cm observation. The orange solid line denotes the minimum path connecting these HMFs, given by the MST algorithm. The synthesized beam is shown in the bottom left corner and the scale bar of 0.01 pc is in the bottom right corner.}
\label{MST}
\end{figure*}

Using high spatial resolution Very Large Array (VLA) observations at 1.3, 2.0, 3.6, and 6.0 cm, \citet{2008AJ....136.1455Z} identified four hypercompact (HC) H{\sc ii} regions, VLA 2a, 2b, 2c, and 2d, located in the MM1 region. Figure~\ref{MST} shows the locations of the four HC H{\sc ii} regions, all of which exhibit strong 1.3 cm continuum emission, but none of them was observed in our 3mm and 1.3mm observations.
Based on the centroid positions of the 1.3 cm emission, we find that VLA 2b is spatially coincident with our HMF 7, while the remaining three HC H{\sc ii} regions show no association with any other HMF. The presence of HC H{\sc ii} emission indicates that a more evolved HMF has already emerged within the MM1 region, highlighting the highly asynchronous and diverse nature of star formation in this region.

\begin{figure*}[ht!]
\centering
\includegraphics[width=\linewidth]{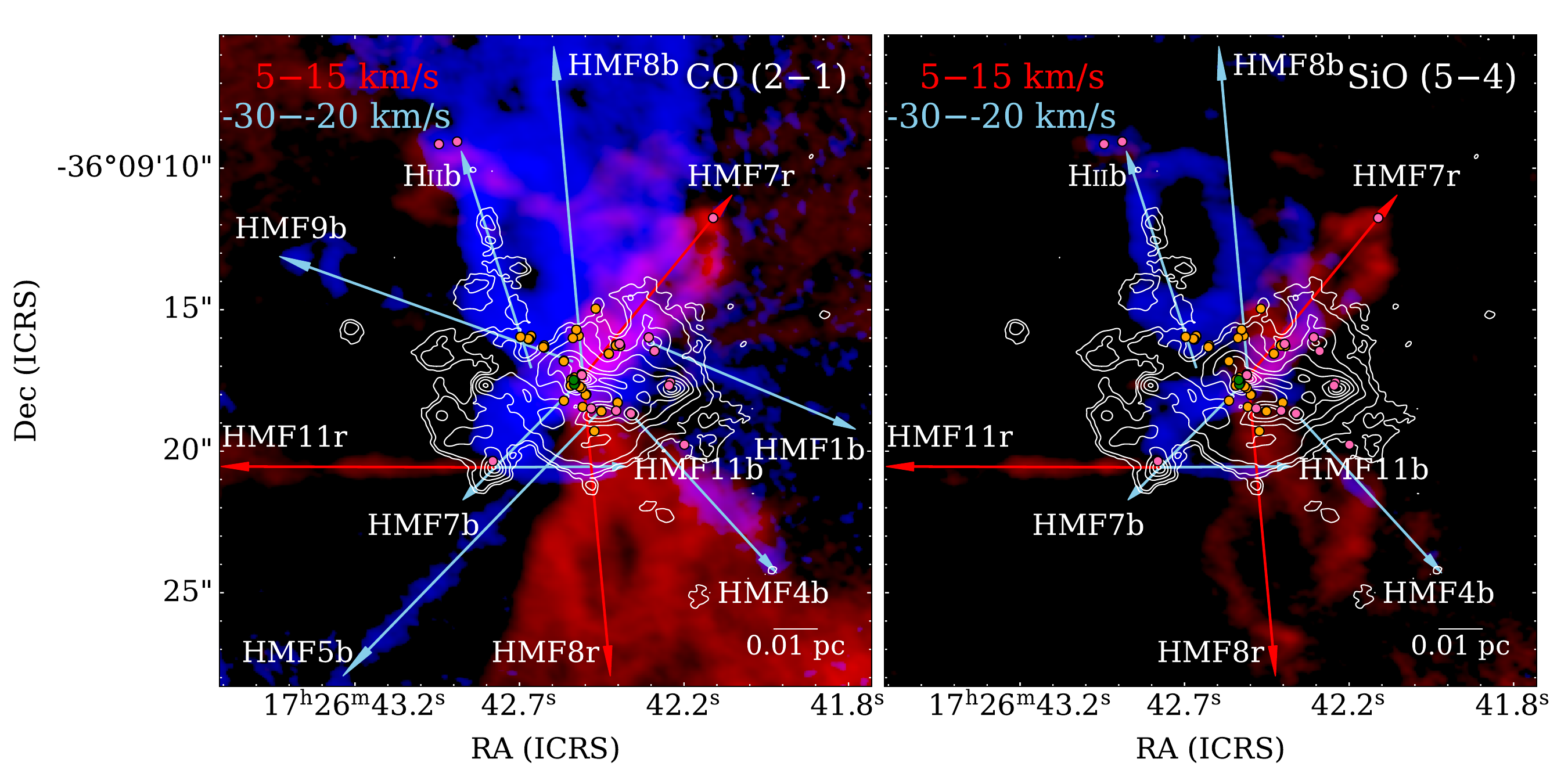}
\caption{Color map of the CO (2–1) and SiO (5–4) toward IRAS 17233-3606 MM1 region, with integrated intensities over 5 to 15 km s$^{-1}$ shown in red and over $-$30 to $-$20 km s$^{-1}$ shown in blue. White contours indicate the 1.3 mm continuum emission. Red and blue arrows denote the red- and blue-shifted molecular outflows, respectively. Class II CH$_3$OH, OH, and H$_2$O masers are marked by green, orange, and pink dots, respectively. }
\label{3color}
\end{figure*}

The complex dynamical activity in IRAS 17233–3606 is characterized by a complex outflow system. We utilized CO (2–1) and SiO (5–4) as complementary tracers to delineate the morphology of these outflows. As shown in the velocity-integrated intensity (moment 0) maps generated by integrating over blue-shifted (-30 to -20 km s$^{-1}$) and red-shifted (+5 to +15 km s$^{-1}$) velocities (see Figure~\ref{3color}), the CO emission reveal a network of 11 distinct outflow lobes. While CO traces the bulk entrained molecular gas, the SiO emission highlights 8 more compact lobes. The SiO emission specifically identifies high-velocity, shock-dominated regions where silicon is sputtered from dust grains \citep{1997A&A...321..293S, 2008A&A...482..809G, 2022A&A...664A..44B}, thereby probing the most energetic feedback components near the protostellar engines. 

Notably, the orientation of the blue-shifted outflow from HMF 8 (HMF8b) is constrained by its close alignment with the bow-shock structures previously identified by \cite{2008AJ....136.1455Z}. For these HMFs where only a single-polarity lobe was detected, the missing counterpart might be attributed to insufficient outflow intensity, and one side is obscured in the dense and complex gas environment within the central MM1 core.
Furthermore, CO (2–1) and SiO (5–4) observations also clearly reveal a pair of north-south distributed outflow cavity walls, which conform to the hourglass structure driven by a protostellar jet \citep{2015ApJ...798L..33H}. 
This jet is likely powered by the central HC H{\sc ii} region VLA2d, given that the outflow gas (labeled as H{\sc ii}b) is well-collimated with the dust continuum and is spatially coincident with the two H$_2$O masers at the tip of the northern cavity wall. 

In IRAS 17233–3606, maser positions were obtained from high-precision catalogs in previous research, including Class II CH$_3$OH masers \citep{1998MNRAS.301..640W}, OH masers \citep{2005ApJS..160..220F}, and H$_2$O masers \citep{2008AJ....136.1455Z}, and overlaid on the ALMA continuum map in Figure~\ref{3color}. Adopting a conservative association radius of $\sim1\farcs0$ ($\sim$1300 au at 1.32 kpc), we found a striking spatial correlation between maser clusters and the HMFs. 
H$_2$O and OH masers are predominantly distributed along the axes of the outflows, consistent with their origin in primary jet-driven shocks \citep{2022A&A...664A..44B, 2024AJ....167...63M}. 
In contrast, Class II CH$_3$OH masers are preferentially located near the center of HMF7, likely excited by the infrared radiation field of the central protostars. 
The spatial correlation between masers and the outflow axes suggests that most of the HMFs are undergoing active accretion,
which in turn drives the enhancement of molecular abundances observed in the gas phase through shock-induced sublimation.

\subsection{Evolutionary Sequence}

\begin{figure}[ht!]
\centering
\includegraphics[width=\linewidth]{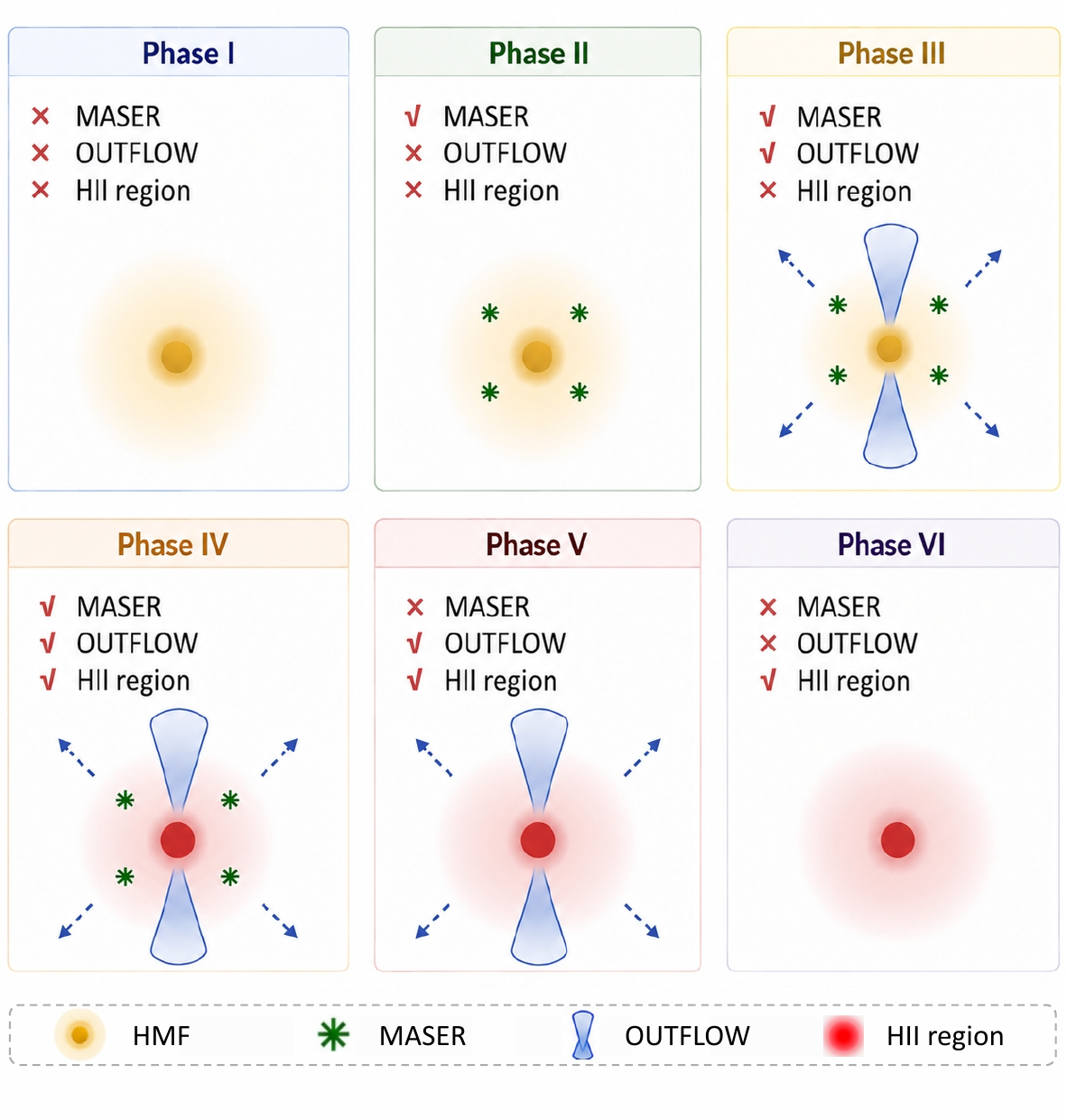}
\caption{Evolutionary sequence of HMFs forulated according to the criteria proposed by \citet{2017A&A...604A..60B}. The six panels demonstrate the chronological development from an embedded protostar (Phase I) to a UC H{\sc ii} region (Phase VI), highlighting the emergence and disappearance of maser emission, outflows, and H{\sc ii} regions.}
\label{phase}
\end{figure}

Following the criteria of \citet{2017A&A...604A..60B}, we used the distributions of masers, outflows, and H{\sc ii} regions to determine the evolutionary sequence for HMFs (see Figure~\ref{phase}).
We first categorize the 11 HMFs in MM1 into four primary evolutionary phases (Phase I–IV), 
based on the distributions of water, OH and CH$_3$OH masers, CO and SiO outflows, and H{\sc ii} region emission. 
Within each evolutionary phase, a relative ranking of the HMFs is tentatively guided by the f[CH$_3$CN/CH$_3$OH] abundance ratio. This method was originally proposed by \cite{2016ApJ...824...99M} based on a limited sample of clumps, serving as a potential chemical clock to indicate the chemical enrichment level in early star-forming regions. However, when observational uncertainties are considered, the ratios in Table~\ref{tab:detection} exhibit overlap and show no clear
monotonic trend across the macro-evolutionary stages (Phases I–IV). This ranking should therefore be served as a tentative indicator of localized chemical enrichment rather than a definitive evolutionary clock. The general validity of this tracer remains to be firmly established through statistically significant studies with larger samples.

It should be noted that, due to projection effects and the highly disordered gas kinematics, it is challenging to unambiguously identify the true driving sources of the outflows and masers. Therefore, the evolutionary sequence presented here should be regarded as tentative. Nevertheless, it is clear that nearly all HMFs are in a relatively early evolutionary stage, with only HMF7 being associated with an H{\sc ii} region.

The coexistence of Phase I–IV HMFs within a small projected area ($\sim$0.1 pc) reveals a significant age spread and non-simultaneous star formation within the MM1 protocluster. While the HMFs of Phase I remain dynamically quiescent, the progression toward Phase III is marked by a clear increase in f[CH$_3$CN/CH$_3$OH] abundance ratio alongside the development of powerful bipolar outflows. This chemical and dynamical synergy indicates that the hot core chemistry is already well-advanced prior to the formation of an H{\sc ii} region. The observed evolutionary gradient suggests that the fragmentation of the massive MM1 region has produced a cluster of protostellar seeds with low initial masses. 

\begin{deluxetable*}{lcllcc}[htp]
\tablewidth{0pt}
\tablecaption{Evolutionary phases of the HMFs.\label{tab:detection}}
\tablehead{
\colhead{ID} & \colhead{Phase} & \colhead{Number and type of maser} & \colhead{Outflow} & \colhead{H{\sc ii}} & \colhead{f[CH$_3$CN/CH$_3$OH]} 
}
\startdata
HMF10 & I & $-$ & $-$ & N & (4.7$\pm$0.9)$\times$10$^{-3}$ \\
HMF6 & I & $-$ & $-$ & N & (1.5$\pm$0.2)$\times$10$^{-2}$ \\
HMF2 & I & $-$ & $-$ & N & (1.6$\pm$0.3)$\times$10$^{-2}$ \\
HMF3 & II & 2$\times$H$_2$O & $-$ & N & (1.4$\pm$0.5)$\times$10$^{-2}$ \\
HMF5 & III & 1$\times$OH & CO & N & (1.1$\pm$0.2)$\times$10$^{-2}$ \\
HMF11 & III & 1$\times$H$_2$O & CO; SiO & N & (1.5$\pm$0.4)$\times$10$^{-2}$ \\
HMF4 & III & 1$\times$H$_2$O & CO; SiO & N & (1.6$\pm$0.3)$\times$10$^{-2}$ \\
HMF1 & III & 2$\times$H$_2$O & CO & N & (2.2$\pm$0.3)$\times$10$^{-2}$ \\
HMF9 & III & 1$\times$OH & CO & N & (2.3$\pm$0.2)$\times$10$^{-2}$ \\
HMF8 & III & 4$\times$H$_2$O & CO; SiO & N & (6.0$\pm$0.8)$\times$10$^{-2}$ \\
HMF7 & IV & 6$\times$OH; 5$\times$CH$_3$OH, class II & CO; SiO & Y & (6.3$\pm$1.7)$\times$10$^{-2}$ \\
\enddata
\tablecomments{The HMFs are arranged in sequence according to their evolutionary phases from younger to more evolved.}
\end{deluxetable*}

\section{Discussion}
\label{dis}

\subsection{Dynamics of MM1: From Fragmentation to Accretion}

To assess the global properties of MM1, we adopt a rotational temperature of T$\rm_{rot}$ $\approx$ 100 K and a FWHM of 4.5 km s$^{-1}$, derived from CH$_3$OCHO in the Band 3 survey \citep{2022MNRAS.511.3463Q}. 
The total velocity dispersion is $\sigma_{tot}$ = $(c_{s}^2 + \sigma^2_{nt})^{1/2}$ $\approx$ 2.0 km s$^{-1}$, 
where the thermal sound speed is c$_s$ = $\rm (k_B T / \mu m_H)^{1/2}$ $\approx$ 0.6 km s$^{-1}$ (with $\mu$=2.33 adopted as the mean molecular weight and T=T$\rm_{rot}$ derived from CH$_3$OCHO) and the non-thermal velocity dispersion is $\sigma_{nt}$ = $\rm(FWHM^2/8ln2 - \rm{k_B T / \mu_{mol} m_H})^{1/2}$ $\approx$ 1.9 km s$^{-1}$ (with $\rm\mu_{mol}$=60 adopted for CH$_3$OCHO), calculated following the method described in \cite{2023MNRAS.520.3259X}.
Given the non-detection of ionized gas emission from the HC H{\sc ii} regions in Band 3, and considering that the total flux density of the 4 HC H{\sc ii} regions from VLA 1.3 cm observations is below 22 mJy beam$^{-1}$, we conclude that their contribution to 3 mm continuum emission is negligible.
Consequently, adopting an average volume density of the Band 3 MM1 core n(H$_2$) $\approx$ 1.7 $\times$ 10$^7$ cm$^{-3}$ from \cite{2025A&A...694A.166C}, 
the calculated turbulent Jeans length is $\lambda_J^{tur}$ = $\sigma_{tot} (\pi / G \rho)^{1/2}$ $\approx$ 1.1$\times$ 10$^4$ au, while the calculated thermal Jeans length is $\lambda_J^{th}$ = $\rm c_{s}(\pi / G \rho)^{1/2}$ $\approx$ 3.3$\times$ 10$^3$ au.

Core separation is a reliable metric for analyzing fragmentation properties \citep{2018A&A...617A.100B}. Here we employed the minimum-spanning tree (MST) method \citep{1985MNRAS.216...17B,2019A&A...629A.135D} to quantify the spatial distribution of the 11 HMFs in MM1 (see Figure~\ref{MST}). 
The projected separations ($\rm d_{MST}$) range from $5.6\times 10^2$ to $3.7\times 10^3$~au, with a mean distance of $\sim$$1.8\times 10^3$~au. Even accounting for the projection effect, the deprojected mean separation remains smaller than the thermal Jeans length ($\lambda_J^{\rm th} \approx 3.3\times 10^3$ au), suggesting that the fragmentation of MM1 is dominated by a thermal Jeans instability. Since the observed mean distance is shorter than $\lambda_J^{\rm th}$, it indicates that MM1 may have experienced gravitational contraction, leading to the closer proximity of the inner HMFs \citep{2024ApJS..270....9X}.

To quantify the spatial distribution of the HMFs, we adopt the $Q$ parameter following the method of \cite{2004MNRAS.348..589C}, defined as $Q = \bar{m} / \bar{s}$. The normalized mean edge length of the MST, $\bar{m}$, is given by $\bar{m} = \sum_{i=1}^{N_c-1} L_i / \sqrt{N_c A}$, where $N_c$ is the number of HMFs, $L_i$ is the length of each edge, and $A = \pi R_{\text{cluster}}^2$ is the cluster area. The normalized correlation length, $\bar{s}$, is defined as $\bar{s} = L_{\text{av}} / R_{\text{cluster}}$, where $L_{\text{av}}$ $\approx$ 4000 au represents the mean separation length between all HMFs. 
As proposed by \citet{2004MNRAS.348..589C}, a value of $Q \gtrsim 0.8$ signifies a centrally condensed spatial distribution, whereas $Q \lesssim 0.8$ indicates a fragmented distribution characterized by subclustering. Our derived value of $Q = 0.77$ suggests that the HMFs are currently in a relatively dispersed state with inherent subclustering. As this value approaches the 0.8 threshold, it likely indicates a transitional phase, where the HMFs are evolving from an initial fractal-like fragmentation toward a more centrally concentrated configuration driven by global gravitation.

\begin{figure*}[ht!]
\centering
\includegraphics[width=0.94\linewidth]{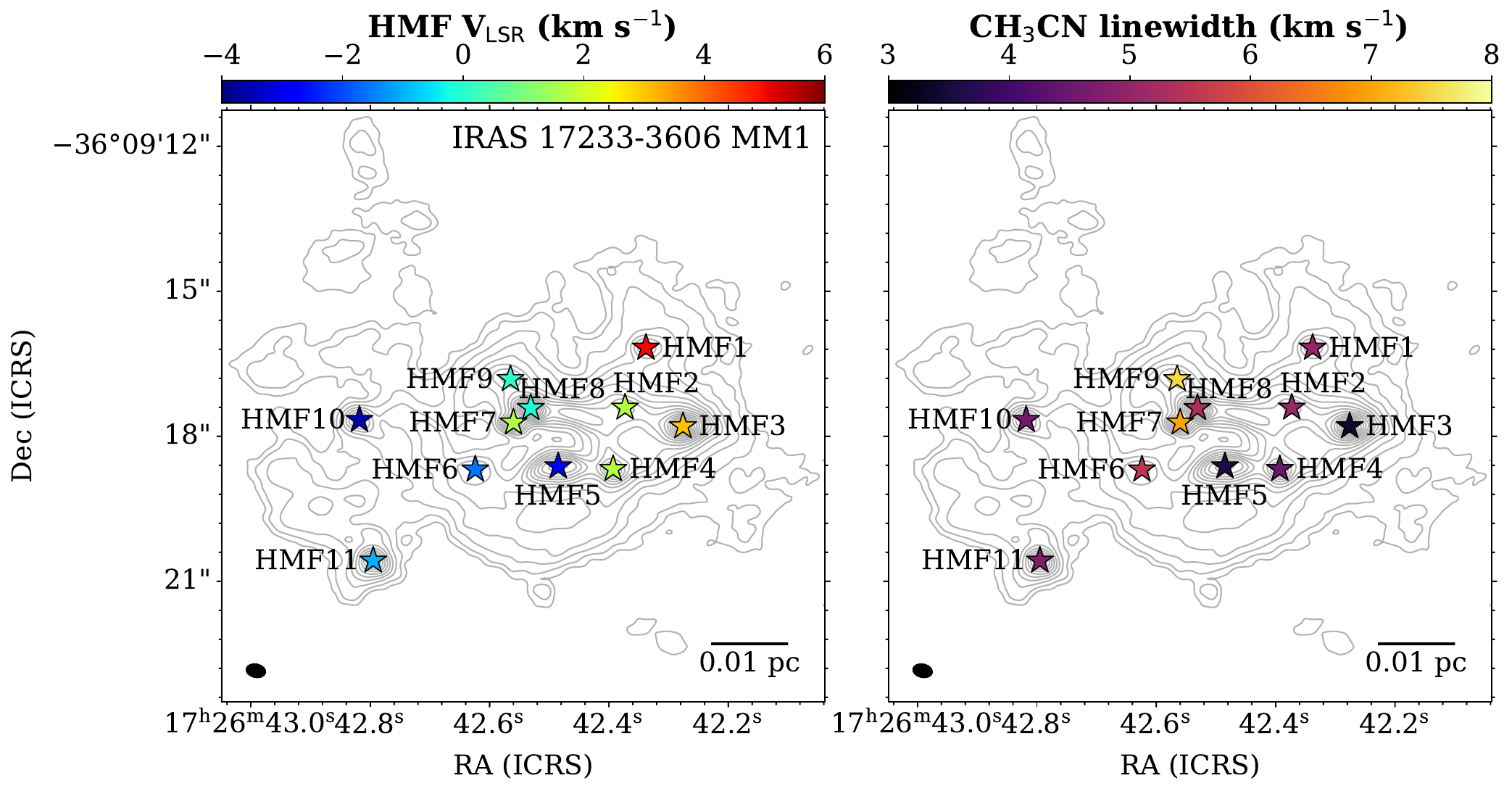}
\caption{HMF V$\rm_{LSR}$ (left) and CH$_3$CN linewidths (right) estimated from the CH$_3$CN fits to the HMFs towards IRAS 17233-3606 MM1. Gray contours are overlaid, following the same power-law progression as Figure~\ref{zoomin}. The stars represent the locations of the HMFs identified by ALMA 1.3 mm observation, with the color scale displaying the fitted parameters from the CH$_3$CN fits (left: core V$\rm_{LSR}$, right: linewidth). The HMF V$\rm_{LSR}$ is the centroid velocity of the CH$_3$CN fit minus the cloud V$\rm_{LSR}$ (taken as -3.2km s$^{-1}$).}
\label{vlsr}
\end{figure*}

The global dynamical condition of the MM1 region was assessed by calculating the virial parameter
$\alpha_{\rm vir}$ = $M_{\rm vir} / M_{\rm gas}$ = 5$\sigma^2_{\rm tot}R_{\rm eff}$/$GM_{\rm{gas}}$. Here, 
the effective radius is $R_{\rm eff}$ = 0.025 pc, and the gas mass is $M_{\rm{gas}}$ = 81.3 M$_\odot$.
The resulting value of $\alpha_{\rm vir}$ = 0.84
is below the critical value for gravitational contraction ($\alpha_{\rm vir}$ $\approx$ 2), indicating that the MM1 region is undergoing global gravitational contraction and actively forming stars.
The result provide an explanation for why the $\rm d_{MST}$ values are shorter than $\lambda_J^{\rm th}$.

In Fig.~\ref{vlsr}, we show the positions of the HMFs together with their respective V$\rm_{LSR}$ and CH$_3$CN linewidths overlaid on the 1.3 mm continuum contour map toward IRAS 17233$-$3606 MM1.
The HMFs exhibit noticeable variations in both systemic velocities and linewidths, indicating highly dynamical gas motions within the central protocluster.
The measured HMF V$\rm_{LSR}$ values span from $\sim-3.6$ km s$^{-1}$ to 5.0 km s$^{-1}$, corresponding to an HMF-to-HMF velocity of $\sim2.5$ km s$^{-1}$.
The HMFs located near the cluster center exhibit V$\rm_{LSR}$ values closer to the systemic velocity, implying that they are more deeply embedded within the central gravitational potential well.
Furthermore, the CH$_3$CN linewidths range from $\sim3.4$ km s$^{-1}$ to 7.5 km s$^{-1}$, with the broadest linewidths preferentially associated with the central HMFs, while relatively narrower linewidths are observed toward several outer sources. 
This trend suggests enhanced dynamical activity toward the cluster center, where the dense gas is likely more strongly influenced by the global gravitational potential. 
\cite{2025A&A...695A..51B} reported mass inflow rates close to $10^{-3}$ M$_\odot$ yr$^{-1}$ along the two filamentary structures connected to MM1 on scales of $\sim0.25$ pc, implying that substantial amounts of material can be continuously transported from the larger-scale environment into the central protocluster, thereby enabling the HMFs within the protocluster to rapidly gain mass through ongoing accretion and eventually form massive stars.

Our analysis suggests that massive star formation in the MM1 region originates from HMFs produced by thermal Jeans fragmentation. Since all HMFs are in Phases I to IV and most exhibit active outflows, a continuous accretion process is evidently ongoing within the region. 
In this situation, the most rapidly evolving cores may grow further by accreting material from the ambient MM1 gas or through potential core coalescence \citep{2007MNRAS.381L..40D, 2023ApJ...959...88D}. These results highlight that the early stages of high-mass star formation are characterized by highly dynamical heterogeneity on scales of $\sim$1000 au.

\subsection{Inner Differentiation in MM2}

\begin{figure*}[ht!]
\centering
\includegraphics[width=\linewidth]{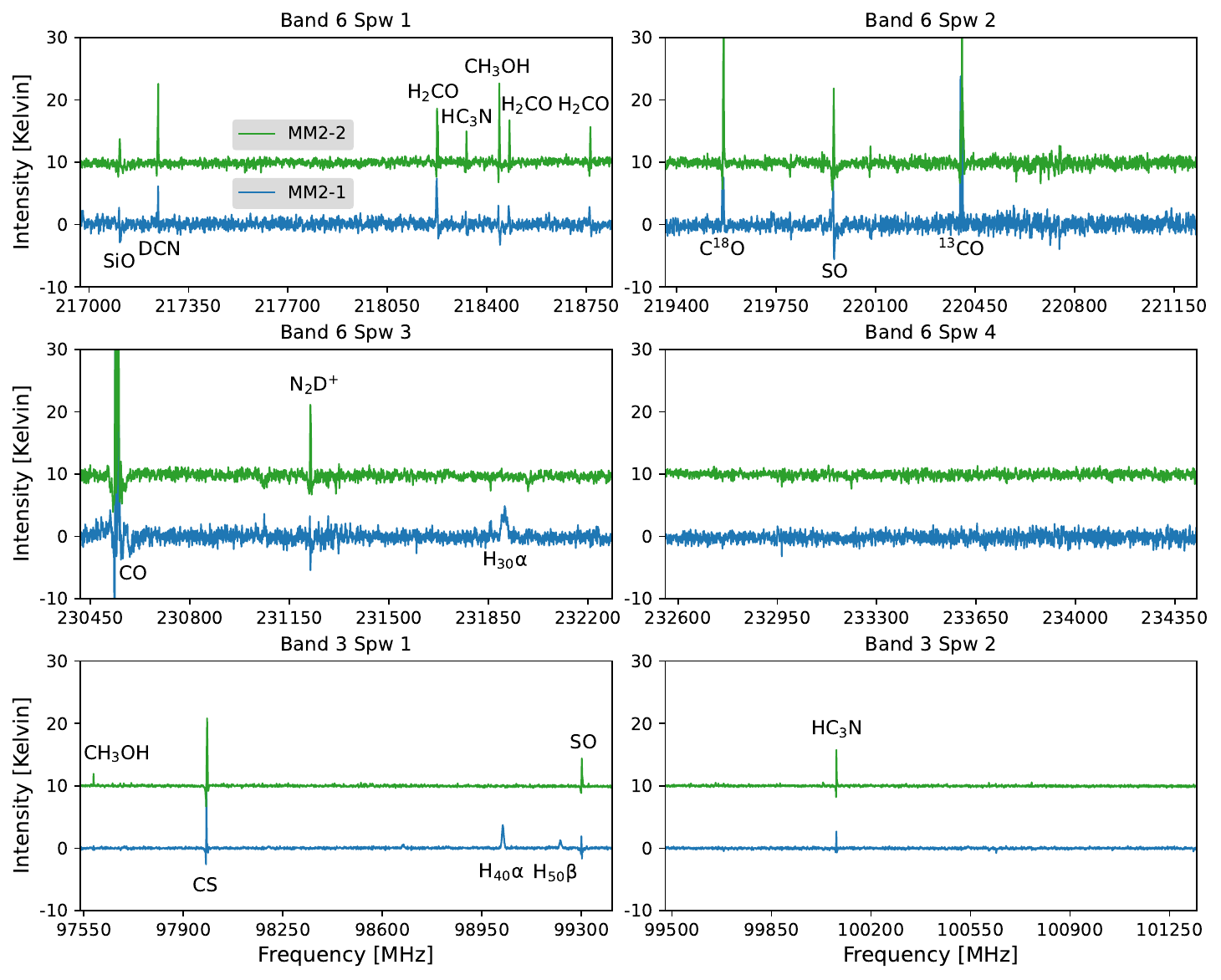}
\caption{Core-averaged spectra of the dense cores MM2-1 (blue) and MM2-2 (green) in the UC H{\sc ii} region. Spectra of MM2-1 and MM2-2 are offset by 10 K in intensity for visual clarity. 
Both cores clearly exhibit a molecular component. However, MM2-1 contains significantly more ionized gas, as evidenced by the detection of H$_{30}$$\alpha$, H$_{40}$$\alpha$ and H$_{50}$$\beta$. In contrast, the detection of CH$_{3}$OH and non-detection of ionized gas suggests that MM2-2 is likely a cold dense core.}
\label{spectra_MM2}
\end{figure*}

The MM2 region exhibits remarkable physical and chemical complexity. There is a significant spatial offset and flux difference between its 3 mm continuum peak of MM2 and the 1.3 mm continuum peak of MM2-1, indicating that this region is likely dominated by an UC H{\sc ii} region ionized by a central massive protostar. The non-detection of this ionizing source at 1.3 mm despite its prominence at 3 mm could be attributed to high dust opacity at shorter wavelengths, which effectively shields the free-free emission \citep{1995ApJ...449..663G}.

Figure~\ref{spectra_MM2} presents the primary molecular transitions detected toward MM2-1 and MM2-2. 
MM2-1, located near the 3 mm continuum center, exhibits multiple radio recombination lines (e.g., H$_{30}\alpha$, H$_{40}\alpha$, and H$_{50}\beta$), which confirm the presence of highly ionized gas consistent with a UC H{\sc ii} region stage. In contrast, despite its relative offset from the primary MM2 peak, MM2-2 lacks any discernible markers of ionized gas and only shows fundamental molecular emission. In particular, the presence of N$_2$D$^+$ indicates that MM2-2 remains in an early stage that is not yet disrupted by ionized gas \citep{2005A&A...439.1023C}. This coexistence of an ionized core and a cold dense core highlights the dramatic localized variations in star formation within the IRAS 17233-3606 complex.

\subsection{Dynamical Feedback from ZAMS star}

\begin{figure*}[ht!]
\centering
\includegraphics[width=\linewidth]{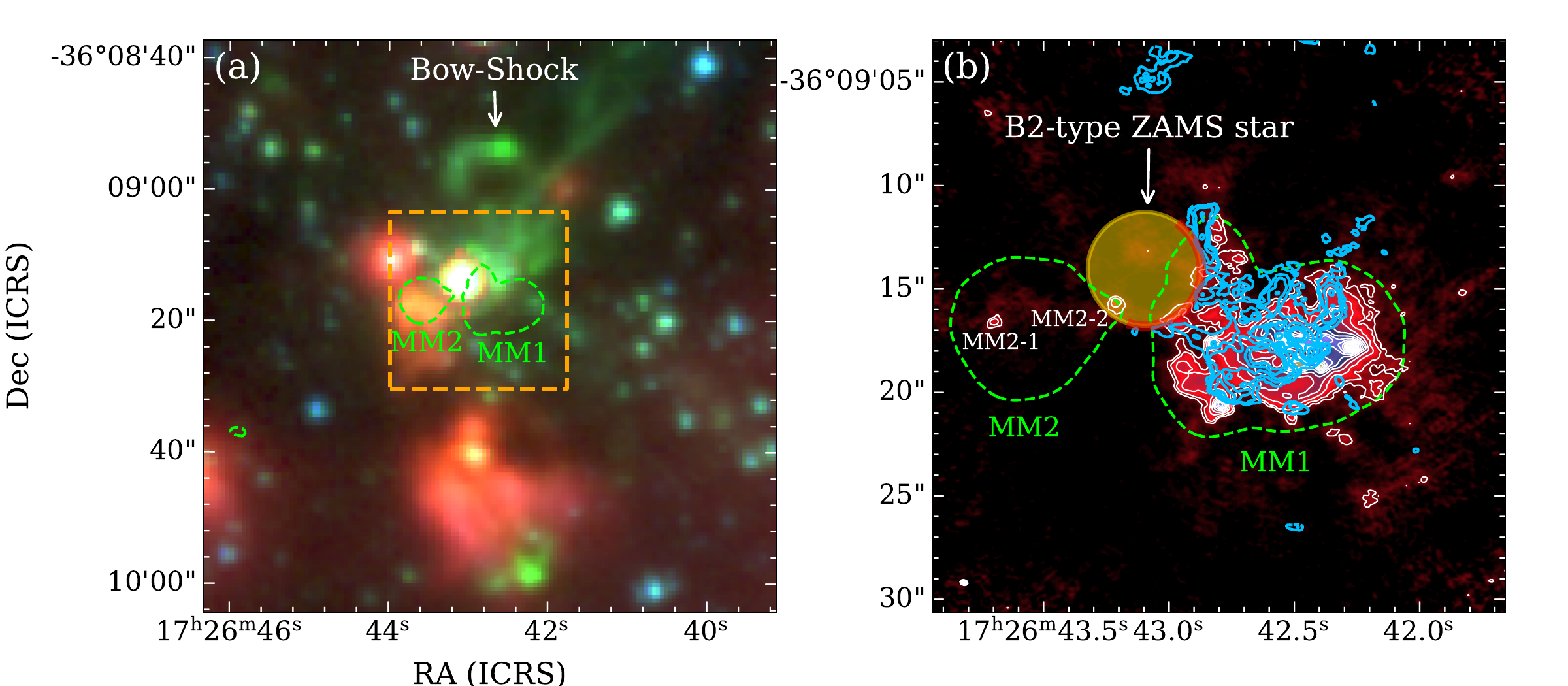}
\caption{Panel (a) shows a three-color mid-infrared image of IRAS 17233–3606 from Spitzer IRAC, with 8.0 µm in red, 4.5 µm in green, and 3.6 µm in blue. The orange dashed square indicates the field of view (FoV) of the ALMA Band 3 observations. The white arrow indicates a bow shock exhibiting prominent emission in the 4.5 µm band. The green contours represent the outermost ALMA Band 3 continuum emission.
Panel (b) shows a zoomed-in view of the ALMA Band 3 FoV. The background image is the ALMA Band 6 continuum emission, overlaid with green outer contours of the Band 3 continuum and white contours of the Band 6 continuum, following the same contour levels as Figure~\ref{zoomin}. The blue contours show the SiO (5–4) moment 0 emission integrated over velocities from –16 to –8 km s$^{-1}$. The yellow circle marks the position of the central B2-type ZAMS star. The red arc highlights the interface of interaction between ZAMS stars and SiO (5-4) gas as well as dust emission.} 
\label{IR}
\end{figure*}

O- and B-type Zero-Age Main-Sequence (ZAMS) stars can exert significant feedback that shapes the formation and evolution of massive stars in the Galaxy \citep{2002ARA&A..40...27C}. As shown in Figure~\ref{IR}, within IRAS 17233-3606 region, the B2-type ZAMS star originally identified by \citet{2008AJ....136.1455Z} drives significant dynamical and radiative feedback onto the adjacent gas of MM1 A prominent arc interface traced by SiO (5-4) emission centered around -12 km s$^{-1}$ and also dust emission indicates a clear interaction between the powerful stellar winds (or outflows) from the B2-type ZAMS star and the dense molecular material of MM1. This morphology strongly suggests that the enhancement of SiO abundance results from the sputtering of grains within shocks, providing direct evidence for the pivotal role of ZAMS stars in shaping the early dynamical evolution of protoclusters. 

An examination of the CH$_3$CN velocity offsets shows that MM1 and MM2 share a similar systemic velocity of approximately $-3.2$ km s$^{-1}$, consistent with their association within the same star-forming complex at a distance of 1.32 kpc. However, in contrast to MM1, though the MM2 and B2-type ZAMS star have overlap areas in the direction of light, there is no clear evidence of a direct interaction between them. One possible explanation is that MM2-2 is a cold dense core, and we cannot directly observe feedback from the ZAMS star since this vicinity lacks prominent outflowing gas. Alternatively, the high density of the MM2 core may provide effective self-shielding \citep{2005MNRAS.358..291D, 2009ApJ...703.1352K}. The dense envelope gas likely intercepted the feedback energy from the ZAMS star and even the ionizing radiation from the central protostar. This allowed the chemically young MM2-2 core to remain undisturbed in a quiescent environment.

\section{Conclusions}
\label{con}

We have investigated the physical and chemical properties of IRAS 17233-3606, using both 3 mm and 1.3 mm ALMA data from the ATOMS and QUARKS survey, respectively. Our observations reveal that IRAS 17233-3606 is a dynamically complex and relatively young environment, providing new observational evidence to current theoretical frameworks. Our major results are as follows:

(1) In the massive hot core MM1 ($\sim$ 81.3 M$_\odot$), we resolved 11 hot molecular fragments (HMFs) that coexist within a projected area of only 0.1 pc and span evolutionary phases as defined by I to IV. The spatial distribution of these HMFs reveals asynchronous star formation within the protocluster.

(2) Since the mean projected separation ($\sim$$1.8\times 10^3$~au) is nearly half of the thermal Jeans length ($\sim3.3\times 10^3$ au), the fragmentation of MM1 is primarily dominated by thermal instability. Meanwhile, the $Q$ parameter $Q = 0.77$ and virial parameter $\alpha_{\rm vir} = 0.84$ both indicate that the MM1 region is undergoing global gravitational contraction and actively forming stars under the influence of gravity.

(3) We detected an interface outlined by SiO (5–4) emission and dust emission in MM1, which spatially couples with the surface of the central B2-type ZAMS star. This directly confirms the impact of stellar feedback on the surrounding medium. Meanwhile, the presence of a cold dense core, MM2-2, within the evolved UC H{\sc ii} region of MM2 suggests that local shielding effects may block external radiation, leading to asynchronous evolution.

\section*{Acknowledgments}
S.-L. Qin is supported by the National Key R$\&$D Program of China (No. 2022YFA1603101) and National Natural Science Foundation of China (NSFC) through grant No. 12433006.
Tie Liu acknowledges the support by the National Key R$\&$D Program of China (No. 2022YFA1603101), National Natural Science Foundation of China (NSFC) through grants No.12073061 and No.12122307, the PIFI program of Chinese Academy of Sciences through grant No. 2025PG0009, and the Tianchi Talent Program of Xinjiang Uygur Autonomous Region.
The research was carried out in part at the Jet Propulsion Laboratory, California Institute of Technology, under a contract with the National Aeronautics and Space Administration (80NM0018D0004).
MJ acknowledges the support of the Research Council of Finland Grant No. 348342.
A.H.  acknowledges the support of the S. N. Bose National Centre for Basic Sciences under the Department of Science and Technology, Government of India, and the CSIR-HRDG, Government of India, for the funding of the fellowship 
LB gratefully acknowledges support by the ANID BASAL project FB210003.
AS gratefully acknowledges support by the ANID BASAL project FB210003.
AS is gratefully supported by the China-Chile Joint Research Fund (CCJRF No. 2312). CCJRF is provided by Chinese Academy of Sciences South America Center for Astronomy (CASSACA) and established by National Astronomical Observatories, Chinese Academy of Sciences (NAOC) and Chilean Astronomy Society (SOCHIAS) to support China-Chile collaborations in astronomy.
This paper makes use of the following ALMA data: ADS/JAO.ALMA\#2021.1.00095.S and 2019.1.00685.S. ALMA is a partnership of ESO (representing its member states), NSF (USA) and NINS (Japan), together with NRC (Canada), NSTC and ASIAA (Taiwan), and KASI (Republic of Korea), in cooperation with the Republic of Chile. The Joint ALMA Observatory is operated by ESO, AUI/NRAO and NAOJ

\vspace{5mm}
\facility{ALMA}

\software{astropy \citep{2013A&A...558A..33A, 2018AJ....156..123A, 2022ApJ...935..167A}, CASA \citep{2007ASPC..376..127M, 2022PASP..134k4501C}, XCLASS \citep{2017A&A...598A...7M}, MAGIX \citep{2013A&A...549A..21M}.}

\bibliography{17233}{}
\bibliographystyle{aasjournal}

\end{document}